\RequirePackage{fix-cm}
\documentclass[twocolumn]{svjour3}          % twocolumn
\smartqed  % flush right qed marks, e.g. at end of proof
\usepackage{graphicx}
\usepackage{xcolor}
\usepackage{lipsum}
\usepackage{multicol}
 \usepackage{mathptmx}      % use Times fonts if available on your TeX system
\usepackage{latexsym}
\begin{document}

\twocolumn[\begin{@twocolumnfalse}

\title{Driver Safety Development: Real-Time Driver Drowsiness Detection System Based on Convolutional Neural Network%\thanks{Grants or other notes
}
\author{ Maryam Hashemi $^{\star}$ \and Alireza Mirrashid \and Aliasghar Beheshti Shirazi }


\institute{Maryam Hashemi \at
	Iran University of Science and Technology, Resalat highway, Tehran, Iran \\
	Tel.: +98913-263-4028\\
	\email{maryam\textunderscore hashemi@elec.iust.ac.ir}           %  \\
	\and
	Alireza Mirrashid \at
	\email{a\textunderscore mirrashid@elec.iust.ac.ir}
	\and
	Aliasghar Beheshti Shirazi \at
	\email{abeheshti@elec.iust.ac.ir}
}

\date{Received: date / Accepted: date}

\maketitle
%\twocolumn[\begin{@twocolumnfalse}
\begin{abstract}
This paper focuses on the challenge of driver safety on the road and presents a novel system for driver drowsiness detection.
In this system, to detect the falling sleep state of the driver as the sign of drowsiness, Convolutional Neural Networks (CNN) are used with regarding the two goals of real-time application, including high accuracy and fastness.
Three networks introduced as a potential network for eye status classification in which one of them is a Fully Designed Neural Network (FD-NN), and others use Transfer Learning in VGG16 and VGG19 with extra designed layers (TL-VGG).
Lack of an available and accurate eye dataset strongly feels in
the area of eye closure detection. Therefore, a new comprehensive dataset proposed.
The experimental results show the high accuracy and low computational complexity of the eye closure estimation and the ability of the proposed framework on drowsiness detection.
\keywords{Drowsiness \and Driver\and Convolutional Neural Networks \and Transfer learning \and Safety }

\end{abstract}.
%\end{@twocolumnfalse}]
\end{@twocolumnfalse}]
	
\section{Introduction}
\let\thefootnote\relax\footnote{Maryam Hashemi}
\let\thefootnote\relax\footnote{ \email {maryam\textunderscore hashemi@elec.iust.ac.ir}}
\let\thefootnote\relax\footnote{\email{maryamhashemi1995@gmail.com}}
\let\thefootnote\relax\footnote{Tel.: +98913-263-4028}

\label{intro}
According to published reports from the World Health Organization (WHO), traffic accidents are one of the top 10 causes that lead to death in the world \cite{1}. The reports demonstrate that the first cause of such crashes are drivers. Therefore, the detection of driver drowsiness could be a suitable methodology to prevent accidents. It also improves the performance of the Advanced Driver Assistance Systems (ADAS) and Driver Monitoring System (DMS); as a result, road safety.

There are three main categories of drowsiness detectors: Vehicle-based \cite{2}, Signal-based \cite{3}, and Facial feature-based \cite{4}.
Vehicle-based methods try to infer drowsiness from vehicle situation and monitor the variations of steering wheel angle, acceleration, lateral position, etc. However, these approaches are too slow for real-time tasks. Signal-based methods infer drowsiness from psychophysiological parameters. Several studies have been done during the last years based on these methods  \cite{5}. The most critical physiological signals that used and investigated are ElectroEncephaloGram (EEG)  \cite{6}, ElectroOculoGram (EOG) \cite{7}, activities of the autonomous nervous system from ElectroCardioGram (ECG) \cite{8}, Skin Temperature (ST), Galvanic Skin Response (GSR) and also intramuscular activities as ElectroMyoGram (EMG). These approaches need to consider invasive captors, which can affect driving negatively. Facial feature-based methods can evaluate the target in real-time without invasive instruments, and they are inexpensive and more accessible than other methods.

\section{Related Work}
Driver behaviors can be extracted to detect driver drowsiness \cite{9}. Deng \textit{et al.} proposed a method for face detection by using landmark points and track the face to find fatigue drivers. They checked signs like yawning, eye closure, and blinking. The system is named DriCare \cite{10}. Zhao \textit{et al.} used both landmark points and texture for face situation classifier. They considered different parts of the face like nose, mouth, and eyes to evaluate the role of every single part of the face for fatigue detection. In the end, they consider the eye and mouth as a dominant sign of fatigue \cite{11}. Verma and colleagues followed \cite{11} strategy and used two VGG16 convolution neural networks parallel for driver emotion detection.
The input of the first network is the detected region of interest (ROI), and in the second VGG16 network, used facial landmark points as input. They involved the results of both networks for deciding \cite{12}. Another work concentrate on bus driver fatigue and drowsiness detection. Based on the bus driver position and window, the eye needs to be examined by an oblique view, so they trained an oblique face detector and an estimated percentage of eyelid closure (PERCLOS) \cite{13}. In \cite{14}, a new dataset for driver drowsiness detection is introduced. They called dataset ULG Multi modality Drowsiness Database (DROZY), and \cite{15} used this dataset with Computer Vision techniques to crop the face from every frame and classify it (within a Deep Learning framework) in two classes: “rested” or “sleep-deprived”. They implemented the system in a low-cost Android device. Another dataset created by Abtahi \textit{et al.} presents two video datasets of drivers with various facial characteristics, like with and without glass and sunglass or different ethnicity for designing algorithms for yawning detection. They also report 60\% accuracy for yawing detection when the camera is located on the dashboard \cite{16}.  The Kinect camera is another instrument for drivers monitoring and identifies driving tasks in a real vehicle. In \cite{17}, authors try to detect seven everyday tasks performed by drivers, normal driving, left-, right-, and rear-mirror checking, mobile phone answering, texting using a mobile phone with one or both hands, and the setup of in-vehicle video devices. The Kinect camera consists of color and depth image information from the driver inside the vehicle. They evaluated 42 features given by Kinect and predicted feature importance using random forests and chose some of them. A feed-forward neural network (FFNN) is used as a learning network. They achieved to 80.7\% accuracy with the FFNN network for task classification.

Eye closeness detection is a challenging task since there are ambient factors that affect it, as lighting condition, image resolution, driver position, different shape of eyes, a different threshold for the closure of an eye, etc. \cite{17}. Kim \textit{et al.} proposed a fuzzy-based method for classifying eye openness and closure. Their approach uses the information of the I and K colors from the HSI and CMYK spaces. Then, the eye region is binarized using the fuzzy logic system based on I and K inputs \cite{18}.  Ghoddoosian \textit{et al.} \cite{19} presented a large and public real-life dataset, which consists of 30 hours of video, with a range of content from subtle signs of drowsiness to more obvious ones. The core of the method is a Hierarchical Multi-scale Long Short-Term Memory (HM-LSTM) network, which is fed by detected blink features in sequence. To confronting noise and scale changes, Song \textit{et al.} \cite{20} proposed a new feature descriptor named Multi-scale Histograms of Principal Oriented Gradients (MultiHPOG) and used feature extraction approach, they tried this method on two different datasets and compared the accuracy and time complexity. Zhao \textit{et al.} used a deep learning method to classify eye states in facial images. The proposed method combines two deep neural networks to construct a deep integrated neural network (DINN). Each network extracts different features and fusion data to make the best decision. Also, a transfer learning strategy is applied to overcome the dataset shortage. This method tested on three datasets and reported 97.20\% accuracy on the ZJU dataset with their DINN network \cite{21}. Another research tried to find the landmark points to find the Eye Aspect Ratio (EAR) and Eye Closure Ratio (ECR) as a sign of drowsiness \cite{22}. Eyelid closure (PERCLOS), blink frequency (BF), and Maximum Closure  Duration  (MCD) used as the features in the Support Vector Machine (SVM) methods \cite{23}. Anitha \textit{et al.} tried to improve the performance of face detection algorithms and track the driver's eye from an input video \cite{24}. Recurrent Neural Networks (RCNN) are prevalent in the driver drowsiness detection field due to they can learn a sequence of features as normal blinking and differs it from falling sleep situation \cite{25}-\cite{28}. Jabbar \textit{et al.} focused on compressing a driver monitoring system based on landmark point that can be implemented on Android application \cite{29}.

This paper aims to propose a drowsy driver warning system that detects the real-time driver’s eye closure. In this system, if the driver's eye classifies as the close class for successive frames, it is a sign of drowsiness, and an alert will be sent to the driver early enough to avoid an accident. 
The proposed work of this article includes four contributions. 1) The frames enter to preprocessing unit to detect the eyes and crop them. Then this unit applies the grayscale function and normalizes the histogram of the eye. The authors used low-resolution images to have real-time detection and histogram equalizer to overcome the bad illumination condition. 2) The authors proposed three neural networks regarding the parameter of fastness, high accuracy, and small dataset. 3) There is a proposed algorithm to evaluates the result of network detection; if it detects as a close eye, the system adds one to the counter, and if the number of the counter reaches more than 12 successive frames, that eye is close, send an alarm for the driver. Otherwise, it keeps the counter for the next frame, whenever an open eye categorizes, counter restart. In other words, the task of this counter is counting successive close frames for distinguishing blinking from falling sleep. 4) Collecting a new dataset that considers a new position of the eye, named oblique view of the driver.
For eye classification to the classes of open and close, three networks considered. The first network is the fully designed neural network, the second is a deep neural network with transfer learning, and it uses a pre-trained VGG16 network, which uses the low-level features on the ImageNet dataset and high-level features to learn. The third network is similar to the second network, but it uses VGG19 with the same goal as the third network. The results show high accuracy and short computational time in the proposed methodology of this article.
The rest of this paper is organized as follows: Section 3 introduces the proposed networks and dataset. Section 4 presents experiments to verify the effectiveness and robustness of the method and dataset. The conclusion discussed in Section 5.

\section{Proposed Work}

In this section, first, the authors peruse the algorithm of eye detection and preprocessing, then the data collection process describes for driver drowsiness detection and reviews another dataset in this field. Finally, the system framework presented in detail.

\subsection{Region of Interest Selection}
\begin{figure*}[h]
	\begin{center}
		\label{1}
		\centering
		\includegraphics[width=17cm,height=5cm]{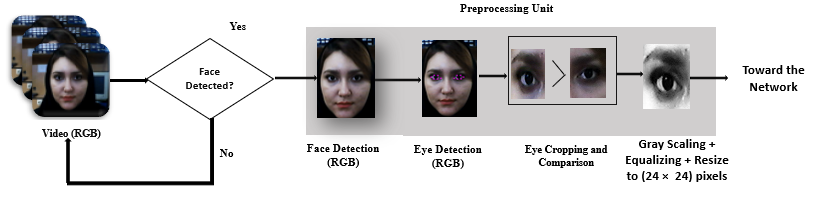}
		\centering
		\caption{Proposed Framework for ROI Selection.}
	\end{center}
\end{figure*}

In this step, we try to find and prepare the eye region and then fed it into the network. The proposed framework of this part is shown in  \figurename {1}. To find the eyes, first need to estimate the headbox, Viola and Jones algorithm \cite{30} used for head detection and facial landmark approach is an implementation of Kazemi and Sullivan \cite{31} work. This is a regression tree machine learning method and was trained on the iBUG 300-W face landmark dataset \cite{32}. The advantage of this approach to others is the high accuracy of detection in different head positions. After reaching the landmark point of the eye, ROI is cropped. The face is asymmetry shape, hence for drowsiness detection, only one eye is adequate to be considered. This approach decreases the computational time for detection noticeably. The eye front of the camera should be selected because it contains more information and decreases the error rate. For this goal, the algorithm computes distances from the most right and the most left point of the right and left eyes and compares them to select the greater distance for cropping. The intended distance is shown in \figurename {2}. The algorithm finds eye landmarks, measure the absolute distance between point 37 and 40, 43 and 46, and, select more significant distance as the eye front of the camera. To overcome the lighting condition's challenge, the authors use a histogram equalizer to equalize eye contrast. This approach makes the methodology more accurate for eye closeness detection. After eye equalizing and detection, the equalized eye sent to the network.

\begin{figure}[h]
	\begin{center}
		\label{2}
		\centering
		\includegraphics[height=7cm,width=7cm]{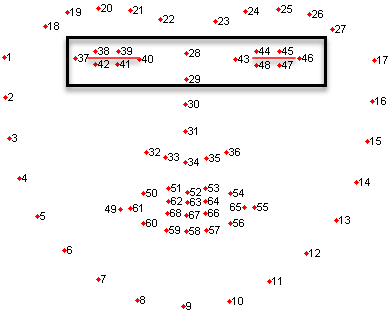}
		\caption{Intended Eye Distance (Point 37 to 40 and 43 to 46) to crop the ROI. }
	\end{center}
\end{figure}
\subsection{Datasets}
Two datasets examined for this research. The first one is the ZJU dataset, and the second one is a mixture dataset of ZJU and the created database of the authors. The details of each described in the following.

%\IEEEpubidadjcol

\subsubsection{ZJU dataset}
The first dataset is the ZJU gallery from the ZJU Eyeblink Database \cite{33}. A set of images gathered from 80 video clips in the blinking video database. They recorded from 20 individuals, four clips for each individual: one clip for frontal view without glasses, one clip with frontal view and wearing thin rim glasses, one clip for frontal view and black frame glasses, and the last clip with an upward view without glasses. Also, images of the left and the right eyes are collected separately. However, since the face is symmetric, the authors use a symmetry-based approach in a further process. It is worth to mention that the use of a subsampled, gray-scale version of the images is enough \cite{34}. In this article,  only the bigger eye used, both right and left eyes trained through the algorithm.  Some samples of the dataset are shown in \figurename {3}.

Dataset has 4841 images, including 2383 closed eyes and 2458 open eyes. All these images are geometrically normalized to the images with 24$\times$24 pixels to train network with sub-sample images. The purpose of training networks with low-resolution images is that the network could be applied to low resolution and illumination states. It worth mentioning, this approach also decreases computational complexity because of limited pixels that should be evaluated.

\begin{figure}[h]
	\begin{center}
		\label{3}
		\centering
		\includegraphics{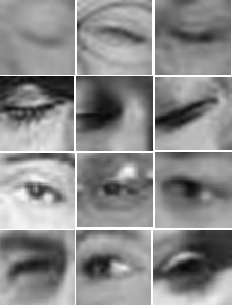}
		\caption{Illustration of Some Closed Left Eye (the First Row), Closed Right Eye (the Second Row), Open Left Eye (the Third Row) and Open Right Eye (the Fourth Row) Samples from ZJU Dataset \cite{34}.}
	\end{center}
\end{figure}

\subsubsection{Proposed dataset}
In this research, the ZJU dataset extended with datasets consists of 4,157 images (2100 open and 2057 close images) from 4 different persons. The authors develop this dataset with a 720p HD webcam camera and collect these images from a video with six frames per second rate. We down-sampled data to the shape of 24$\times$24, to be comparable with ZJU data and can be used in a lousy illumination environment as night time. The strength point of this dataset is considering the different situations of the eyes. The data collected from different distances rotates and angles, which increases the driver’s freedom of action. Data are also subtended drivers with and without glasses. Ring of glasses divided into thick and thin rings. Data categorized into two groups. The first category considers the data that the driver’s head looks straight with different eye rotations. The second category applies the data the driver turns his head, but in the permissible range, it means the camera recorded side view. A second category is a novel approach for eye dataset, which containing oblique view.
The network of this article is trained with both straight and oblique view, simultaneously. Also, ZJU contains only Chinese ethnicity; however, the data here contains another ethnicity, which expands a variety of train data. The proposed methodology of this article involves different lighting conditions, and all of the images pass through a histogram equalizer. Some eye image samples of looking forward and turning head positions are shown in \figurename {4} and \figurename {5}, respectively. Displayed images are after equalizing the brightness. The extended dataset is combined with ZJU to a comprehensive database to train the model.
\begin{figure}[h]
	\begin{center}
		\label{4}
		\centering
		\includegraphics[width=8cm,height=3cm]{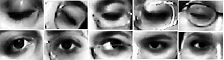}
		\caption{Display of Straight Head for Close Eye (the First Row) and Open Eye (the Second Row) from Extended Dataset.}
	\end{center}
\end{figure}

\begin{figure}[h]
	\begin{center}
		\label{5}
		\centering
		\includegraphics[width=8cm,height=3cm]{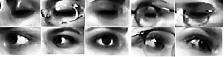}
		\caption{Display of Spin Head for Close Eye (the First Row) and Open Eye (the Second Row) from Extended Dataset.}
	\end{center}
\end{figure}

\subsection{Network Architecture}
As mentioned earlier, three different neural networks proposed in this work, and the result of these networks compared with each other for both datasets. The first network is a Fully Designed Neural Network (FD-NN). The architecture and layers of the model are displayed in \tablename{ \ref{one}}. A 2D convolutional layer with 3$\times$3 filter size used, and Relu assigned as an activation function. Maxpooling with the size of 2$\times$2 applied to reduce the number of features. 25\% of connections deactivate in the period of learning to avoid overfitting. Flatten implement to transform data to vector for the next layers. As the last activation layer, we used the Sigmoid function for binary classification output. The selected hyperparameter for each network also mentioned in their section. The supremacy of FD-NN compare to other proposed networks is non-complexity and quickness for real-time tasks. This network applied to ZJU and extended dataset.

\begin{table}
	\centering
	\caption{Proposed Fully Designed Neural Network (FD-NN) Structure.}
	\label{one}
	\begin{center}
		\begin{tabular}{| p{3cm} | p{2cm}| p{1.5cm} |p{1.5cm}|}
			\hline
			\centering
		Layer Name & Size of Feature Map & Number of filters\\ \hline
			\centering
			Image input layer & 24 $\times$24 $\times$3 & --- \\ 
			\centering
			Conv2D & 32 $\times$ 32 $\times$ 32 & 32  \\ 
			\centering
			Maxpooling2D & 32 $\times$ 31 $\times$ 30 & 1 \\ 
			\centering
			Dropout & 32 $\times$ 31 $\times$ 30 & 1\\ 
			\centering
			Fully Connected Layer & 1 $\times$ 1 $\times$ 512 & 1 \\ 
			\centering
			Sigmoid & 2 &---  \\ \hline
			
		\end{tabular}
	\end{center}
\end{table}

In the second and third networks, the concept of transfer learning and pre-trained CNN used to extract features. For this purpose, VGG16 and VGG19  selected as the proposed pre-trained networks. VGG16 is a deep convolutional neural network proposed by  Simonyan and Zisserman \cite{35}. VGG16 contains 16 layers and trained on the ImageNet dataset, which consists of a large number of images and having 1000 classes. VGG19 is a more in-depth version of VGG16 with 19 layers. It also trained on the ImageNet database. The network designed for images with the size of 224$\times$224 pixels, nevertheless, can imply other sizes, either. These networks learn low-level features with the weight of the ImageNet dataset, and high-level features extract with three last added fully connected layers. The first layer is an input layer, second is activation with Relu function, and the last layer is a Sigmoid function as an output layer. We call these networks Transfer Learning VGG16 (TL-VGG16) and Transfer Learning VGG19 (TL-VGG19). The structure is similar in both networks, and they are different in the term of VGG16 or VGG19.
The framework for these networks presented in \figurename {6}. The goal of using transfer learning is to have a deeper network with higher accuracy, especially when the training dataset is small, and we are not able to train a deep neural network with that dataset. Transfer learning is also able to reduce training time.

\begin{figure*}[h]
	\begin{center}
		
		\centering
		\label{6}
		\centering
		\includegraphics[height=3cm ,width=15cm]{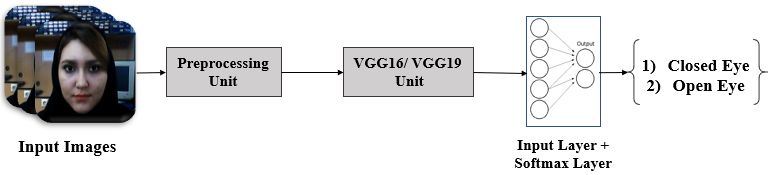}
		\caption{Structure of Transfer Learning VGG16 (TL-VGG16) and Transfer Learning VGG19 (TL-VGG19).}
        
	\end{center}
\end{figure*}

\subsection{Decision Making}
As the last step of detection, images were taken in 6 frames per second. The structure of closure eye detection displayed in \figurename {7}. If the network estimates the probability of closeness of eye more than 50\% for more than 12 successive images, a danger consider and a voice alarm will be sent to the driver. 
Experimental results showed that the average time for blinking is 100ms-400ms for a healthy person. Therefore, closed eyes that last more than 1 second is a symptom of tiredness. To reduce the false-negative result of our algorithm, we decided to announce the alarm if the eyes are closed continuously in 2 seconds. Since we processed six frames per second, 12 successive frames of closed eye detection are drowsiness danger announce.
\begin{figure*}[h]
	\begin{center}
		\centering
		\label{7}
		\centering
		\includegraphics[height=6cm, width=14cm]{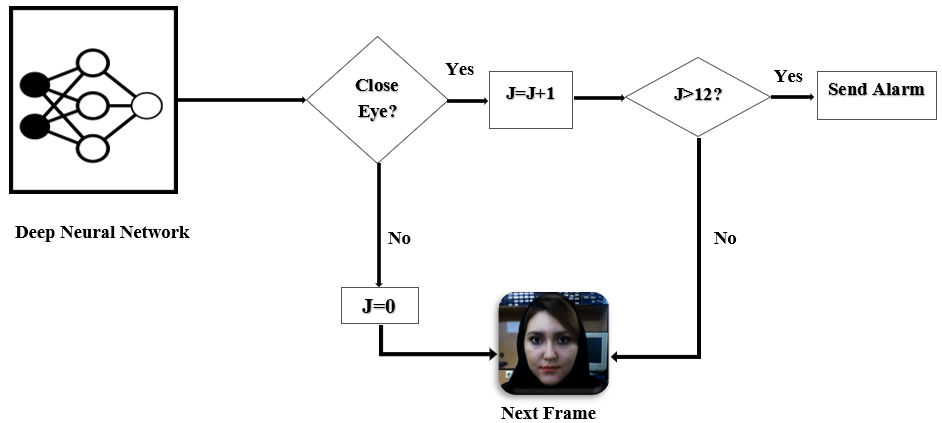}
		\caption{Structure of Eye Closure Detection.}
	\end{center}
\end{figure*}

\section{Results and Discussion}
In this research, we used a system with Intel Core i7-6700K CPU@ 4.00GHz with 16GB RAM and NVIDIA GeForce GTX 1070Ti GPU. The Python selected as language programming on the Anaconda platform. The ZJU dataset \cite{33} and our extended dataset are applied to detect drowsiness. 

\subsection{Accuracy Evaluation}
The ZJU dataset \cite{33}, which is used to test the model, consists of static images from the eye in varying illumination conditions, different pupil directions, and various eye features. The eye in the training process classified into two categories (open and closed regardless it is right or left eye). This ROI, which is selected by landmark points, consider as input to the driver drowsiness detection system using the TL-VGG16 network, TL-VGG19 network, and also the FD-NN. 
The authors trained three networks with the ZJU dataset. The number of images in ZJU datasets is in \tablename  {2}.

\begin{table}
	\caption{Size of ZJU Dataset.}
	\label{2}
	\begin{center}
		\begin{tabular}{|p{2cm}|p{2cm}|p{2cm}|}
			\hline
			\centering
			Total Number & Closed eye & Open eye  \\ \hline
			4157 & 2100 &2057 \\ \hline
			
		\end{tabular}
	\end{center}
\end{table}

In the ZJU dataset, as previous work in this field only used train and validation data; hence, to be able to compare our networks with other proposed works only used train and validation data. Stochastic gradient descent and cross-entropy chose as optimizer and loss function.  0.01 selected as learning rate and 70\% data used for training and rest for validation. The accuracy and Area Under Curve (AUC) of networks on the ZJU dataset shown in \tablename {3}, indicated that the FD-NN, TL-VGG16, and TL-VGG19 gives 98.15\%, 95.45\% and 94.96\% accuracy on ROI images, respectively. The numbers indicate that the selected convolutional neural network method and design of the networks can appropriately satisfy the parameter of accuracy.  As it can see in \tablename {3}, the highest accuracy belongs to FD-NN. In \tablename {4}, the True Positive (TP), False Positive (FP), True Negative (TN), False Negative (FN), recall, and precision of FD-NN are reported to have a comprehensive evaluation of this network. It can be seen that precision in FD-NN is 99.8\%, and recall is 86.7\%.  

\begin{table}
	\caption{Accuracy and AUC of Three Proposed Network on ZJU Dataset.}
	\label{3}
	\begin{center}
		\begin{tabular}{ | p{2.4cm} | p{1.5cm}|p{1.5cm}| p{1.1cm}|}
			\hline
			\centering
			Network & Accuracy & AUC  & Epoch\\ \hline
			\centering
			TL-VGG19 & 94.96\% &99.0\% & 50 \\ \hline
			\centering
			TL-VGG16& 95.45\% &99.0\% & 100\\ \hline
			\centering
			FD-NN & 98.15\%&99.8\% & 50\\ 
			\hline
		\end{tabular}
	\end{center}
\end{table}

\begin{table}
	\caption{Precision and Recall in FD-NN for ZJU Dataset.}
	\label{4}
	\begin{center}
		\begin{tabular}{|p{1.5cm}|p{0.5cm}|p{0.5cm}|p{0.5cm}|p{0.5cm}|p{1cm}|p{0.75cm}|}
			\hline
			\centering
			Total Test Number & TP & FP & TN & FN & Precision & Recall  \\ \hline
			\centering
			1247 & 558 & 1 & 603 & 85 & 99.8\% & 86.7\%\\ \hline

		\end{tabular}
	\end{center}
\end{table}
For each presented network, we repeat the training ten times and report the average accuracy. In \figurename {8}, we displayed the accuracy value for each training time for the FD-NN in ZJU dataset. The variance and standard deviation for reported numbers are 0.231 and 0.480, respectively.

\begin{figure}[h]
	\begin{center}
		\centering
		\label{8}
		\centering
		\includegraphics [height=4cm,width=8cm]{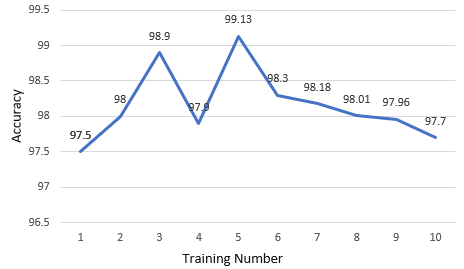}
		\caption{Accuracy Values for Each Training Time in FD-NN for ZJU Dataset.}
	\end{center}
\end{figure}

The authors also experiment on the extended dataset that contains images of ZJU and the proposed images. The proposed images have four categories, including 1) Closed eye and look ahead, 2) Open eye and look ahead, 3) Closed eye and turned head, and 4) Open eye and turned head. The number of images in each category recorded in \tablename {5}. The dataset extracted to the three networks with considering stochastic gradient descent optimizer and 0.01 as the learning rate, and the loss function is cross-entropy. 80\% of data used for training, 10\% validation, and 10\% test.

It can be seen from the \tablename {6} that for the extended training data, the FD-NN, TL-VGG19, and TL-VGG16 networks achieve to 97.01\%, 96.42\%, and 98.53\% accuracy, respectively in the validation part and 96.79\%, 96.09\%, and 97.54\% accuracy, respectively in test part. As mentioned, the highest accuracy belongs to TL-VGG 16. The precision and recall of TL-VGG16 reported in \tablename {7}. The exclusivity of the extended dataset is in containing an oblique view. It is tough to compare three proposed network's results of the extended dataset with only the ZJU dataset because, in our extended dataset, we used the test and validation data, nevertheless in ZJU, only validation result id reported.

\begin{table}
	\caption{Number of Images in Extended Dataset.}
	\label{5}
	\begin{center}
		\begin{tabular}{|p{1cm}|p{1.25cm}|p{1.25cm}|p{1.25cm}|p{1.25cm}|}
			\hline
			\centering
			Total Number & Closed Eye and Look Ahead & Open Eye and Look Ahead & Closed Eye and Turned Head & Open Eye and Turned Head \\ \hline
			4185 & 1521 & 1445 & 558 & 661 \\ \hline
			
		\end{tabular}
	\end{center}
\end{table}

\begin{table}
	\caption{Accuracy and AUC of 3 Proposed Networks on Extended Dataset.}
	\label{6}
	\begin{center}
		\begin{tabular}{ | p{1.5cm} | p{1cm}| p{1cm} |p{1cm} |p{1cm}|p{0.75cm}|}
			\hline
			\centering
			Network & Accuracy of Validation & AUC of Validation & Accuracy of Test & AUC of Test & Epoch\\ \hline
			\centering
			TL-VGG19 & 96.42\% & 99.4\%& 96.09\%& 99.3\% & 100\\ \hline
			\centering
			FD-DNN & 97.01\% & 99.4\%& 96.79\% & 99.3\%&50 \\ \hline
			\centering
			TL-VGG16 & 98.53\% & 99.8\%& 97.54\%&99.5\%  & 100\\ 
			\hline
		\end{tabular}
	\end{center}
\end{table}
\begin{table}
	\caption{Precision and Recall in TL-VGG16 for Extended Dataset.}
	\label{7}
	\begin{center}
		\begin{tabular}{|p{1.5cm}|p{0.5cm}|p{0.5cm}|p{0.5cm}|p{0.5cm}|p{1.25cm}|p{0.75cm}|}
			\hline
			\centering
			Total Test Number & TP & FP & TN & FN & Precision & Recall  \\ \hline
			\centering
			834 & 426 & 5 & 395 & 8 & 98.8\% & 98.15\%\\ \hline
			
		\end{tabular}
	\end{center}
\end{table}
As mentioned in the previous sections, faces are symmetric. Therefore, only the process on one eye serves the purpose of reducing the computation time and the chances of false detections. It can be found from the results of this section that the accuracy of detection in the ZJU dataset with FD-NN is higher than other networks. In \tablename {8}, accuracy and AUC of each method presented in detail. It can be seen that the proposed FD-NN reaches to the highest accuracy and AUC among all of the proposed networks.

\begin{figure}[h]

	\begin{center}
		\centering
		\label{9}
		\centering
		\includegraphics[height=5cm,width=8cm]{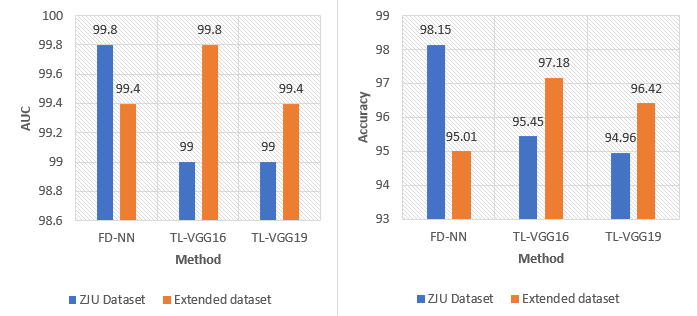}
		\caption{Comparison of Accuracy and AUC of Extended Dataset and ZJU Dataset with Different Networks.}
	\end{center}

\end{figure}

The results of the extended dataset and ZJU dataset also compared in \figurename {9}. The figure illustrates that the extended dataset in VGG16 and VGG19 leads to better accuracy and AUC because the size of the dataset is more extensive than ZJU and is more suitable for deep learning algorithms. It can be concluded the transfer learning decreases the size of the needed dataset, but there is a trade-off about minimum size and accuracy in deep neural networks.

\begin{table}
	\caption{Comparison of Accuracy and AUC among the Proposed Networks and Other Methods on the ZJU Dataset.}
	\label{8}
	\begin{center}
		\begin{tabular}{   p{1.4cm}   p{3.6cm}  p{1cm} p{0.75cm}  }
			\hline
			\hline
			\centering
			Research   & Method   & Accuracy (\%)  & AUC (\%) \\ \hline
			
			Pan \cite{34} & cas-Adaboost (W=3) & 88.8&-\\
			Song \cite{20} & HPOG+LTP+Gabor(s) & 95.91&89.22\\
			Song \cite{20} & MultiHPOG+LTP+Gabor & 96.83& 99.27\\
			Song \cite{20} & MultiHPOG+LTP+Gabor(s) & 96.40& 96.67\\
			Zhao \cite{21} & DNN & 94.45& 97.91\\
			Zhao \cite{21} & DCNN & 95.79&98.15\\
			Zhao \cite{21} & DINN & 97.20 &99.29\\
			Ours  & TL-VGG19 &\textbf{95.00} &\textbf{99.00} \\
			Ours & TL-VGG16 & \textbf{95.45} &\textbf{99.00}\\
			Ours & FD-NN & \textbf{98.15}&\textbf{99.80} \\ \hline
		\end{tabular}
	\end{center}
\end{table}

The advantage of our proposed network over  Pan \cite{34} and Song \cite{20} is in using convolutional neural networks instead of other feature extraction approaches. Indeed, FD-NN tries to find non-linear correlations through convolutional layers. Hence, this network can learn more elaborate features.
In terms of comparing FD-NN to Zhao \cite{21}, and TL-VGG19, TL-VGG16 it worth mentioning that there is a relationship between the number of parameters that the network needs to learn and the size of the dataset. As much as the number of hidden layer increase, the network needs greater dataset. To be more specific, the ZJU dataset's size is more consistent with the number of FD-NN parameters. Hence, the network can update and finally learn its weight more accurately. In other words, the mentioned networks in comparison with FD-NN are too deep for this dataset. Based on the above argument, we expect that performance of TL-VGG16 and TL-VGG19 improve when the size of the dataset extends. The results of \tablename {6} prove this prediction.

\subsection{Complexity Evaluation}
\begin{figure}[h]
	\begin{center}
		\centering
		\label{10}
		\centering
		\includegraphics [height=6cm,width=8cm]{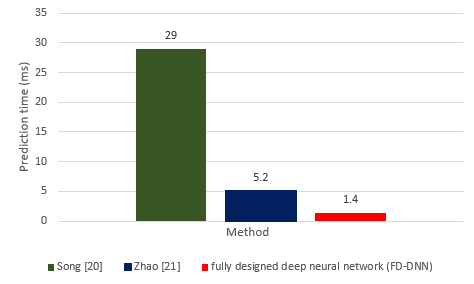}
		\caption{Comparison of Time Cost for Eye Closure Detection in Different Methods on ZJU Dataset.}
	\end{center}
\end{figure}

For a better comparison, the computational complexity should be considered as a significant parameter between the different methods. In real-time applications, faster methods are more reliable. In \figurename {10}, the necessary time for deciding on eye closure for each method on the ZJU dataset presented. The FD-NN is almost 4 times and 20 times faster than the models of \textit{Zhao et al.} \cite{21} and \textit{Song et al.} \cite{20}, respectively. The reason is that the FD-NN uses fewer parameters than the networks in \cite{21} and is a less complicated network than \cite{21}. In comparison with \cite{20}, we use neural networks instead of feature extraction methods that cause to better accuracy and higher speed. Specifications of FD-NN make the decision faster, which helps it to privilege for real-time tasks while it has better accuracy and AUC.

The computational complexity of the ZJU dataset and the extended dataset are the same. As a matter of fact, the computational complexity depends on network architecture than the dataset. The size of images in the ZJU dataset is the same as images in our proposed dataset. It means extended dataset images pixel is the same as the ZJU dataset. Hence, the number of pixels that the network needs to process is equal. In conclusion, the computational complexity of FD-NN in the ZJU dataset is equal to the computational complexity of FD-NN in the extended dataset.

\subsection{Limitation and Implementation}
The authors referred to the benefits of proposed approaches as high accuracy and low computational complexity. However, there is a trade-off between the size of the dataset and accuracy. Deeper networks with more hidden layers and parameters show better performance and need a bigger dataset.  Therefore, there is a compromise between the number of available data and the number of FD-NN parameters.
	
Furthermore, there is a great variety of visual parameters that should be considered during data collection. Lighting conditions, the ethnicity of driver, angel, and location of the camera, humidity, the eye color, and glasses are a few cases. Considering all of the variables is not possible, and the authors selected the most significant parameters. Nevertheless, eliminating other parameters affect the result in the real situation.	

In addition, from a psychological perspective, falling sleep has different steps and cannot be divided into only two categories. Considering sleep as a binary situation means an approximation of intermediate levels that affect performance in the real situation.
Artificial neural network implementation requires high-performance hardware. FPGA and GPU are the most known hardware platforms for neural network implementations. To select a platform, there are many parameters like power, efficiency, and ease of use to consider. The authors think that many predesigned hardwares with GPU such as smartphone and tablets might be used for implementation because there are suitable platforms for implementing artificial neural networks on on-board GPUs.

\section{Conclusion and Future Works}
The drowsiness detection plays a vital role in safe driving, and this paper proposed a new system to prevent accidents arising from drowsiness. Therefore, in the first step of the system, the system captures a stream of frames, and in the preprocessing unit, landmark points applied to access ROI, then eye region is selected, and the preprocessing unit selects the eye front of the camera. Then, the image converts to gray image, and contrast of eye equalizes. Finally, the image resizes to 24$\times$24 pixels. The authors use the output of this step as input to the network to eye state classification. If the network detects the eye is closed for more than 12 successive images, an alarm will be sent for the driver. Otherwise, it considers it as blinking.
The authors also gathered a dataset for driver drowsiness detection, which contains a new state of the eye, named oblique view.  The expended dataset advantage is considering the oblique view, which makes the system works in more varied situations. 
The authors also proposed three networks to achieve better accuracy and less computational time for drowsiness detection based on eye state.
Three networks presented as potential network and accuracy and other parameters evaluated to find the most reliable network. The experimental results indicated that the FD-NN network reaches to 98.15\% accuracy and 99.8\% AUC. FD-NN is also reliable for real-time tasks; the needed time for eye state classification is 1.4 ms in this network.

For future works, the authors intend to focus on yawning analysis for fatigue detection by landmark points of mouth.
Also, a different level of drowsiness can investigate without limitations of being in two situations because the boundaries among different levels of drowsiness are so narrow, and it would be a more challenging task  \cite{36}.

%\begin{acknowledgements}
%\end{acknowledgements}

\section*{Conflict of Interest}
The authors declare that they have no conflict of interest.



\end{document}